\documentclass[twocolumn,showpacs,preprintnumbers,amsmath,amssymb,floatfix]
              {revtex4}
\usepackage{graphicx}              % Include figure files
\usepackage{epsfig}                % Include figure files
\usepackage{dcolumn}               % Align table columns on decimal point
\usepackage{bm}                    % bold math

\newcommand{\be}{\begin{equation}}
\newcommand{\ee}{\end{equation}}

\begin{document}

\preprint{\fbox{\sc version of \today}}

\def\lt{\raisebox{0.2ex}{$<$}}
\def\gt{\raisebox{0.2ex}{$>$}}

\title{A Particle number conserving shell-correction method\thanks{This
       work has been partly supported by the Polish Committee for Scientific
       Research under Contract No. 2P03B~115~19 and by the Program  No. 99-95
       of Scientific Exchange between the IN$_2$P$_3$, France, and Polish
       Nuclear Research Institutions.}}
\author{              K.~Pomorski        }
\affiliation{     Katedra Fizyki Teoretycznej,
            Uniwersytet Marii Curie-Sk{\l }odowskiej,
             PL-20031 Lublin, Poland}

%\date{}

\begin{abstract}

The shell correction method is revisited. Contrary to the traditional 
Strutinsky method, the shell energy is evaluated by an averaging over the 
number of particles and not over the single-particle energies, which is 
more consistent with the definition of the macroscopic energy. In addition, 
the smooth background is subtracted before averaging the sum of 
single-particle energies, which significantly improves the plateau condition 
and allows to apply the method also for nuclei close to the proton or neutron 
drip lines. A significant difference between the shell correction energy 
obtained with the traditional and the new method is found in particular for 
highly degenerated single-particle spectra (as i.e. in magic nuclei) while 
for deformed nuclei (where the degeneracy is lifted to a large extent) both 
estimates are close, except in the region of super or hyper-deformed states.

\end{abstract}


\maketitle

%%%%%%%%%%%%%%%%%%%%%%%%%%%%%%%%%%%%%%%%%%%%%%%%%%%%%%%%%%%%%%%%%%%%%%%%%%%%%%%%
%%%%%%%%%%%%%%%%%%%%%%%%%%%%%%%%%%%%%%%%%%%%%%%%%%%%%%%%%%%%%%%%%%%%%%%%%%%%%%%%
%%%%%%%%%%%%%%%%%%%%%%%%%%%%%%%%%%%%%%%%%%%%%%%%%%%%%%%%%%%%%%%%%%%%%%%%%%%%%%%%

\section{Introduction}
\label{Sec01}

The macroscopic-microscopic method of evaluating the potential energy
surfaces and binding energies of nuclei was proposed in the papers of
Strutinsky \cite{St66} and Myers and \'Swi{\c a}tecki \cite{MS66}. Despite 
the tremendous progress of selfconsistent models to nuclear structure the
macroscopic-microscopic method remains one of the most important tools.
In such an approach the microscopic energy corrections are added to the 
macroscopic part of the nuclear binding energy described by the liquid drop 
model or other macroscopic methods. The microscopic part consists of shell 
and pairing energies. The prescription for the evaluation of the shell energy 
by smoothing the single-particle energy spectra was first given in Ref.\ 
\cite{St66} and than improved in Refs.~\cite{NT69,BD72}. This Strutinsky 
method of averaging over single-particle energies is still widely used up to 
now, in spite of its known problems which appear for mean-field potentials 
of finite depth as well as for nuclei close to the proton or neutron drip 
lines.

Already in the 70' (see Refs.~\cite{SI75,SI77,IS78,IS79,St79,Iv84} and related
papers) Strutinsky and Ivanyuk made an attempt to replace the original
Strutinsky method of evaluating the smooth energy component by an averaging in
the space of particle numbers (${\cal N}$-space) that should be more consistent
with the macroscopic part of the binding energy which is usually evaluated in a
liquid-drop type approach. The parameters of such macroscopic models are
usually obtained by a least-square fit to nuclear masses which corresponds  to
an averaging in the ${\cal N}$-space (e.g. in Ref.~\cite{PD03}). In
Refs.~\cite{SI75,IS78} the smooth component of the total single-particle energy
was approximated by a polynomial in ${\cal N}$-space with coefficients that
were determined by a least-square fit. It was shown in Refs.~\cite{IS79,Iv84}
that the shell correction energies  obtained by these two types of averaging
procedures are not the same. Significant differences appear for highly
degenerated single-particle spectra, as e.g. in spherical nuclei. The method by
Ivanyuk and Strutinsky of finding the smooth energy developed in
Refs.~\cite{SI75,IS78,IS79,Iv84}, has reached sufficient accuracy to be used in
practical calculations. It was, however, never widely used, probably because of
its complexity.

Another way of separating out the smooth part of the sum of single-particle 
energies can be found in Ref.~\cite{SG77}, where the liquid-drop type 
asymptotic expansion of the total single-particle energy in powers
$A^{1/3}$ was used. Unfortunately this method of evaluating the average
energy was not precise enough to be used in practice.

In the present paper a different method of evaluating the shell energy is
proposed. The smooth component of the total single-particle energy is obtained
by folding the sum of single-particle energies in the ${\cal N}$-space 
with a modified Gauss function as described in the Appendix. In addition, an 
average energy background as obtained by the harmonic oscillator energy sum rule
(see section \ref{HO} below) is subtracted before performing the folding, 
which significantly increases the precision of the method. Our new prescription
for the shell correction energy gives results close to those obtained in the 
Ivanyuk and Strutinsky approach of Refs.\ \cite{IS79,Iv84} and is extremely 
simple to use. 

One should also mention that the shell energy evaluated with the present model 
conserves exactly the given number of particles, and not only on the average, 
as was the case in the traditional Strutinsky method.

%%%%%%%%%%%%%%%%%%%%%%%%%%%%%%%%%%%%%%%%%%%%%%%%%%%%%%%%%%%%%%%%%%%%%%%%%%%%%%%%
%%%%%%%%%%%%%%%%%%%%%%%%%%%%%%%%%%%%%%%%%%%%%%%%%%%%%%%%%%%%%%%%%%%%%%%%%%%%%%%%
%%%%%%%%%%%%%%%%%%%%%%%%%%%%%%%%%%%%%%%%%%%%%%%%%%%%%%%%%%%%%%%%%%%%%%%%%%%%%%%%

\section{Theoretical model}
\label{SecTh}

In the macroscopic-microscopic method of evaluating potential energy
one decomposes the nuclear binding energy into three parts
\be
\begin{tabular}{ll}
$E(Z,A;\,{\rm def})$&$= E_{\rm mac}(Z,A;\,{\rm def})
                      + E_{\rm shell}(Z,A;\,{\rm def})$\\
                    &$+ E_{\rm pair}(Z,A;\,{\rm def}) \,\,,$
\end{tabular}
\label{Emac}
\ee
where $Z$ and $A$ are the charge and mass numbers respectively. The macroscopic
part, $E_{\rm mac}$, depends on the deformation of nucleus and is usually
evaluated in the liquid drop or some other more sophisticated model. The
microscopic part of the energy consists of the shell and the pairing energies.
The pairing energy $E_{\rm pair}$ is usually evaluated in the (projected or
not) BCS formalism (see e.g. \cite{NT69} or \cite{BD72}), while the shell
energy $E_{\rm shell}$ is the sum of the  proton and neutron contributions
\be
E_{\rm shell}(Z,A;{\rm def}) = E_{\rm shell}^p(Z;{\rm def}) 
                             + E_{\rm shell}^n(A-Z;{\rm def}) \,\,.
\label{Esh1}
\ee
The shell correction energy of one kind of particles is equal to the 
difference
\be
E_{\rm shell} = \sum\limits_{i=1}^{\cal N} e_i - \widetilde E({\cal N}) \,\,,
\label{Esh2}
\ee
where ${\cal N}$ is the number of particle in the system and $\widetilde E$ is
the smooth part of the total single-particle energy, where {\it smooth} means 
slowly varying with the particle number ${\cal N}$. 
In the following two different methods of evaluating of this smooth part will 
be presented.

%%%%%%%%%%%%%%%%%%%%%%%%%%%%%%%%%%%%%%%%%%%%%%%%%%%%%%%%%%%%%%%%%%%%%%%%%%%%%%

\subsection{Harmonic oscillator energy sum rule}
\label{HO}

The eigenenergies of the spherical harmonic oscillator
\be
e_n = (n+\frac{3}{2})\, \hbar\omega_0
\label{ho1}
\ee
are strongly degenerated
\be
{\rm deg}_n = \frac{1}{2}(n+1)(n+2)\times 2 \,\,.
\label{deg1}
\ee
Here $n=0, 1, 2, ...$ is the main quantum number and $\omega_0$ is the
harmonic oscillator frequency. The factor 2 in the above equation is due to
the two possible orientation of the spin.

According to Ref.~\cite{BM69} the degeneracy of the main harmonic oscillator 
shell can be approximated by
\be
{\rm deg}_n \approx \left(n+\frac{3}{2}\right)^2 
             = \left(\frac{e_n}{\hbar\omega_0}\right)^2 \,\,.
\label{deg2}
\ee
The total number of particles ${\cal N}$ occupying all shells up to 
$n={\rm N}$ is
\be
{\cal N}({\rm N}) = \sum\limits_{n=0}^{\rm N} {\rm deg}_n = \frac{1}{3}
          {\rm (N+1)(N+2)(N+3)}\,\,.
\label{deg3}
\ee
It is easy to show \cite{BM69} that for large ${\rm N}$ values the following
approximation holds:
\be
{\cal N}({\rm N}) \approx \frac{1}{3}\left({\rm N}+\frac{3}{2}\right)^3 
            = \frac{1}{3}\left(\frac{e_{\rm N}}{\hbar\omega_0}\right)^3 \,\,. 
\label{deg4}
\ee
The last equation can serve as the average relation between the single-particle
energy $e$ and the number of particles which occupy the levels with energy 
smaller or equal to $e$
\be
{\cal N}(e) = \frac{1}{3}\left(\frac{e}{\hbar\omega_0}\right)^3 \,\,,
\label{Ne}
\ee
or
\be
  e({\cal N}) = (3{\cal N})^{1/3} \hbar\omega_0 \,\,.
\label{eN}
\ee
Eq.~(\ref{Ne}) leads to the known expression for the average density of the 
harmonic oscillator single-particle levels
\be
g = \frac{\partial{\cal N}}{\partial e} = \frac{e^2}{(\hbar\omega_0)^3}
  = \frac{(3\,{\cal N})^{2/3}}{\hbar\omega_0} \,\,.
\label{den}
\ee
The sum $E$ of single-particle energies of all occupied levels is
\be
E = \sum\limits_{n=0}^{\rm N} e_n\, {\rm deg}_n = \hbar\omega_0 
    \sum\limits_{n=0}^{\rm N} (n+\frac{3}{2})\,(n+1)\,(n+2)
\label{sume}
\ee
and can be approximated by the integral
\be
\overline{E} \approx \int\limits_0^{\cal N} e({\cal N}')\, d{\cal N}' \,\,.
\ee
Inserting here Eq.~(\ref{eN}) one obtains the following energy sum rule:
\be
\overline{E} \equiv \overline{\left(\sum\limits_{i=1}^{\cal N} e_i \right)}
                \approx \frac{1}{4}(3{\cal N})^{4/3} \hbar\omega_0 \,\,.
\label{SHO}
\ee
{\it The sum of energies of nucleons which occupy the harmonic oscillator 
levels is thus proportional to the $4/3$ power of the total number of particles
in the system.}\\ 
A more accurate estimate than the above one was made in Ref.~\cite{BM69} :
\be
\overline{E} \approx \left[\frac{1}{4}(3{\cal N})^{4/3} +
                     \frac{1}{8}(3{\cal N})^{2/3}\right]\hbar\omega_0 \,\,.
\label{SHO1}
\ee
The term proportional to ${\cal N}^{2/3}$ is important in the light systems
but in the heavier nuclei it can be neglected as much smaller than the leading
${\cal N}^{4/3}$ term.
\begin{figure*}
%\begin{widetext}
  \begin{center}
  \includegraphics[height=17cm, angle=270]{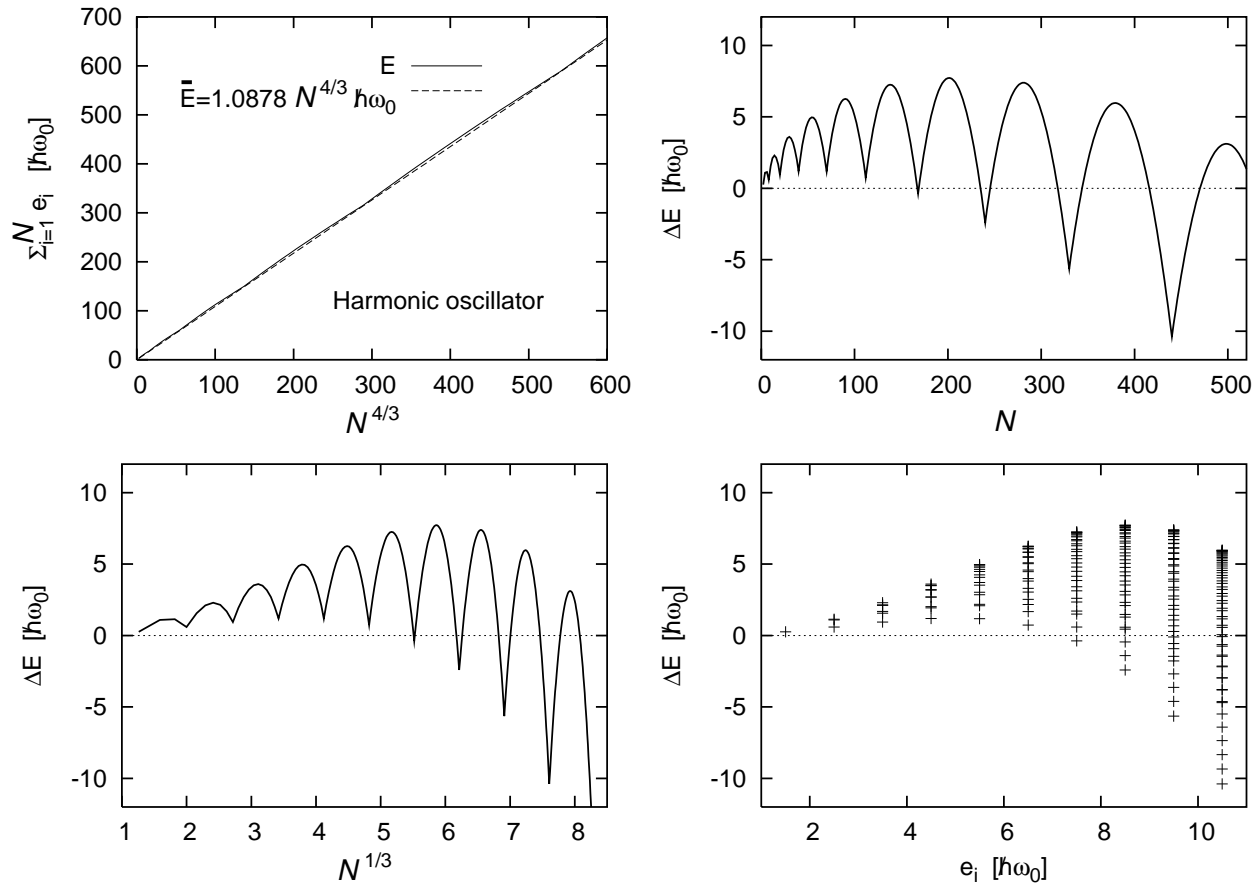}
\end{center}
\caption{Sum of the single-particle energies ($E$, solid line) and its
approximation  ($\overline{E}$, dashed line) by Eq.~(\ref{SHO}) (top l.h.s.
plot) as well as their difference ($\Delta E=E-\overline{E}$) in function of
the number of  particles ${\cal N}$ (top r.h.s. plot), ${\cal N}^{1/3}$ (bottom
l.h.s. plot) and the single particle energies $(e_i)$ (bottom r.h.s. plot).}
\label{fig1}
%\end{widetext}
%\hfill
\end{figure*}

In the top l.h.s.\ part of Fig.~\ref{fig1} the sum $E$ of single-particle
energies (solid line) and its approximation $\bar E$ (dashed line) by
Eq.~(\ref{SHO}) are shown as function of the number of particles ${\cal N}$.
The deviation $\Delta E$ between both lines is hardly visible on this
scale, so we present it separately in the top r.h.s. part of Fig.~\ref{fig1}. 
The coefficient in front of the term ${\cal N}^{4/3}$ was obtained by a least 
square fit and turns out to be very close to the value of the approximate 
expression (\ref{SHO}) which is exact in the limit ${\rm N}\rightarrow\infty$. 
A strong shell structure corresponding to the harmonic oscillator magic 
numbers: ${\cal N}_n$= 2, 8, 20, 40, 70, 112, 168, 240, 330, 440, .... is 
observed. Using Eq.~(\ref{Ne}) one can obtain the average distance between 
the harmonic oscillator major shells as function of the particle number 
\be
{\cal N}_{n+1}^{1/3} - {\cal N}_n^{1/3} = \frac{1}{3^{1/3}}\frac{e_{n+1}-e_n}
    {\hbar\omega_0} = 3^{-1/3} \,\,.
\label{dmag}
\ee
The deviation $\Delta E$ of the energy sum from its average behavior as 
function of ${\cal N}^{1/3}$ is presented in the bottom l.h.s. part of 
Fig.~\ref{fig1}. It is seen that the distance between closed shells is nearly 
constant and roughly equal to $\Delta({\cal N}^{1/3})\approx 0.7$ which is 
the estimate, Eq.\ (\ref{dmag}). It is worth noticing that the same data 
plotted as function of the single-particle energies $e$ shows a structure 
(bottom r.h.s.\ of Fig.~\ref{fig1}) which seems hard to interpret at first 
sight. Obviously the shell structure of the harmonic oscillator is more
visible when one plots $\Delta E$ as function of ${\cal N}^{1/3}$.

The relation (\ref{SHO}) was obtained assuming that the single-particle 
energies are measured with respect to energy zero. Assuming that the minimum 
of the harmonic oscillator potential corresponds to 
$V_0$ (i.e. $e_i \rightarrow e_i+V_0$) one can get the more general relation 
\be
\overline{E}  = \overline{\left(\sum\limits_{i=1}^{\cal N} e_i\right) }\,
                \approx \,a\, {\cal N}^{4/3} + V_0 \,{\cal N} \,\,.
\label{SE}
\ee

We have verified (numerically) that the above harmonic oscillator energy sum 
rule is universal and not only fullfiled by the  spectra of the modified 
harmonic oscillator (Nilsson potential) or other finite depth model mean-field 
potentials (e.g. Saxon-Woods) but also by the single-particle spectra obtained 
selfconsistently for the Hamiltonians associated with the Gogny or Skyrme 
effective forces. 
\begin{figure*}
  \begin{center}
  \includegraphics[height=17cm, angle=270]{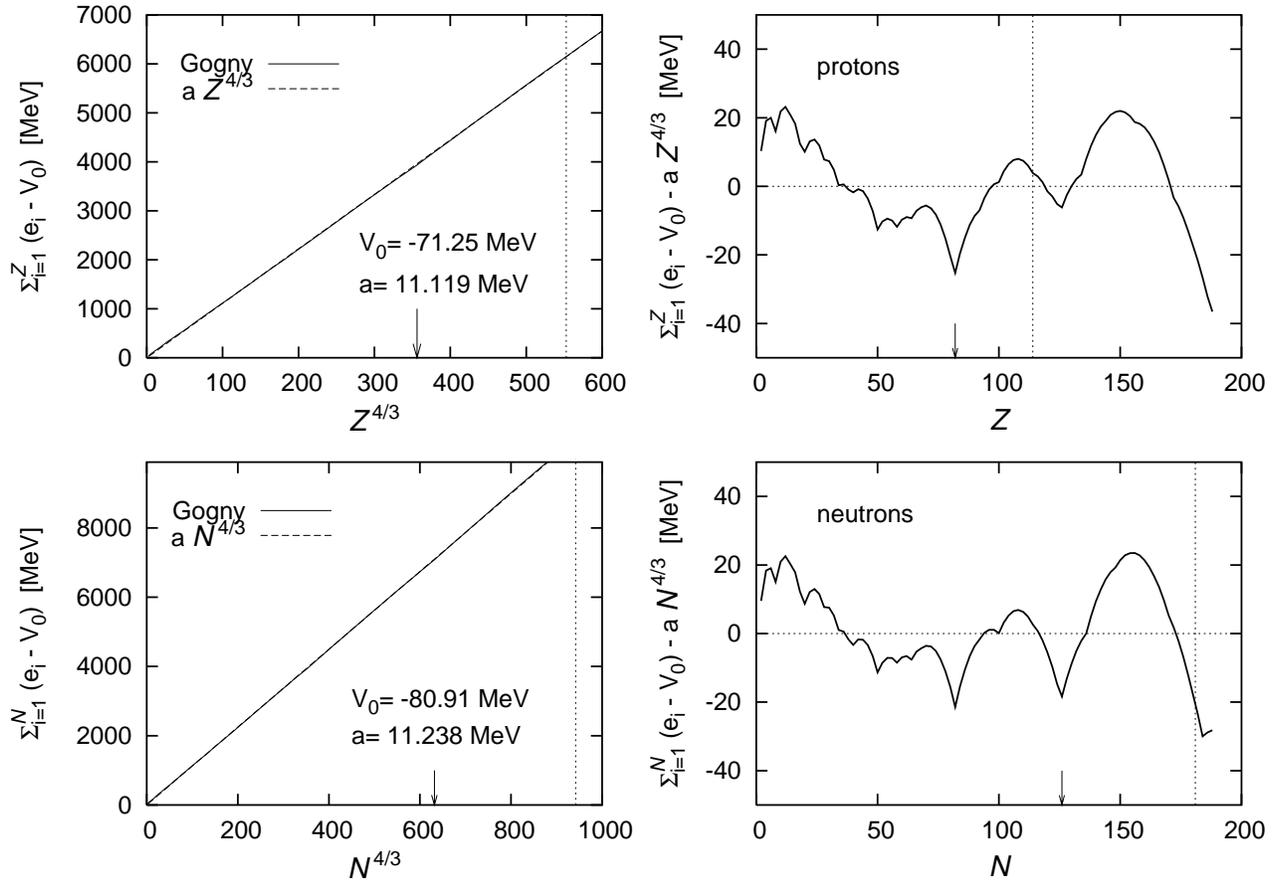}
  \end{center}
\caption{Sum of the single-particle energies ($E$, solid line) obtained
selfconsistently with the Gogny D1S force for $^{208}$Pb and its approximation 
($\overline{E}$, dashed line) by Eq.~(\ref{SE}) as well as the deviation from
this average trend (r.h.s. plots) as function of the number of protons (top) or
neutrons (bottom). Arrows indicate the position of the Fermi energy and the 
vertical lines mark the end of the bound spectrum.}
\label{fig2}
\end{figure*}

A typical deviation of the sum of the single-particle energies (with
respect to the bottom of the effective mean-field potential) from the
estimate (\ref{SE}) is of the order of a few promilles for heavier nuclei. In
Fig.~\ref{fig2} the sum of the single-particle energies (l.h.s. column) and its
deviation ($\Delta E$) (r.h.s. column) from the average trend, Eq.~(\ref{SE}), 
is plotted as function of $Z^{4/3}$ for protons (top row) and $N^{4/3}$ for
neutrons (bottom row). The single particle energies of spherical $^{208}$Pb
were obtained selfconsistently using the Hartree-Fock approximation to the
Gogny Hamiltonian with the D1S force \cite{BG84}. The parameters $a$ and $V_0$
of Eq.~(\ref{SE}) given in Fig.~\ref{fig2} are obtained by a least square
fit. The arrows point the Fermi-level positions and the dotted vertical
lines mark the end of the bound state spectrum. A very pronounced shell
structure of the proton and neutron spectra is visible in the r.h.s. plots.

%%%%%%%%%%%%%%%%%%%%%%%%%%%%%%%%%%%%%%%%%%%%%%%%%%%%%%%%%%%%%%%%%%%%%%%%%%%%%%%%

\subsection{Average of the sum of single-particle energies}
\label{new}

Let us define a discrete sample of data, $S_n$, as the difference between the
sum of the lowest available single-particle energies of the $n$ fermion system 
and the corresponding background energy, $\overline{E}(n)$, obtained using the
harmonic oscillator sum rule, Eq.\ (\ref{SE})
\be
S_n \equiv \sum\limits_{i=1}^n e_i - \overline{E}(n) 
    = \sum\limits_{i=1}^n e_i -a \, n^{4/3} - V_0 \, n \,\,.
\label{Sn}
\ee
The parameters $a$ and $V_0$ are determined by minimizing the square 
deviation between the single-particle energy sum and $\overline{E}$
\be
\sum\limits_{n=1}^{{\cal N}_{max}} S_n^2 = {\rm min}  \,\,,
\ee
where ${\cal N}_{max}$ can be chosen as the maximal number of nucleons which 
can be put on the given single-particle energy spectrum.

Using the Gauss-Hermite folding procedure described in details in Appendix A 
one can evaluate the average value of $S_n$ corresponding to ${\cal N}$ 
nucleons
\be
\begin{tabular}{ll}
$\widetilde S_{\cal N}$ &$= \frac{1}{\gamma\sqrt\pi}
    \sum\limits_{n=2,4}^{{\cal N}_{max}} \frac{2}{3\, n^{2/3}}\, S_n \,
    \exp\left\{-\left(\frac{{\cal N}^{1/3}-n^{1/3}}{\gamma}\right)^2 \right\}$\\
      &$\cdot \,\,f_6\left(\frac{{\cal N}^{1/3}-n^{1/3}}{\gamma}\right) \,\,,$
\end{tabular}
\label{Stilde}
\ee
where $f_6$ is the 6$^{th}$ order polynomial given by Eq.~(\ref{fcorr}). 
The folding is performed not directly in the particle number $n$ but in its 
cubic root since the distance between the major harmonic
oscillator shells is constant in $n^{1/3}$ and approximately equal 0.7 as we
have shown above.
The factor $3n^{2/3}$ in the denominator of Eq.~(\ref{Stilde}) is the direct
consequence of the transformation  $n \rightarrow n^{1/3}$, while the factor 2 
in the numerator is due to the spin degeneracy of the single-particle levels. 

The smoothed energy of an even or odd ${\cal N}$ system is then 
\be
\widetilde E({\cal N}) = \widetilde S_{\cal N} + a {\cal N}^{4/3} + V_0 {\cal N} \,\,
\,\,,
\label{Etilde}
\ee
where we have restored the background energy $\overline{E}({\cal N})$, 
Eq.\ (\ref{SE}), which has been subtracted from the single-particle energy 
sum in Eq.\ ({\ref{Sn}). 
Subtracting $\overline{E}(n)$ in (\ref{Sn}) increases significantly the
accuracy of evaluating the smoothed part $\widetilde E$ of the energy since the
deviations $S_n$ is 2 to 3 orders of magnitude smaller than the value of
$\overline{E}({\cal N})$. The smoothed energy obtained in this way is less
sensitive to the energy cut-off of the single-particle spectrum, which is
important for evaluating the shell energy of nuclei close to the
proton or neutron drip lines.
\begin{figure*}
  \begin{center}
  \includegraphics[height=17cm, angle=270]{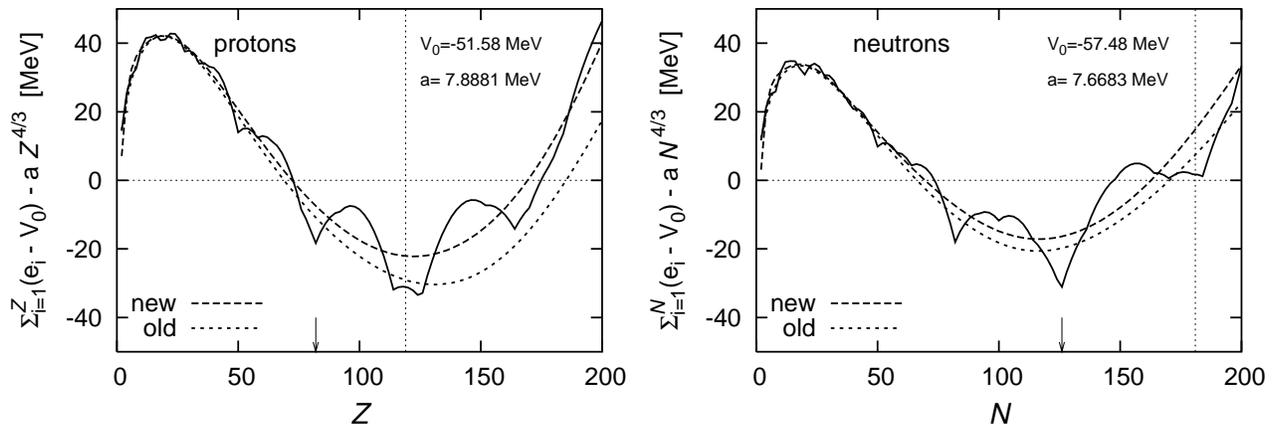}
  \end{center}
\caption{Sum of the single-particle energies $E$ (solid lines) obtained
for the Saxon-Woods potential of $^{208}$Pb and its smooth part obtained
with Eq.~(\ref{Etilde}) (fat dashed lines) as well as within the traditional
Strutinsky method (thin dotted lines). From all the curves is subtracted the
background energy evaluated as in Eq.~(\ref{SE}). The data for protons and for
neutrons are presented in the left and right parts respectively. Arrows 
indicate the positions of the Fermi energies and the vertical dotted lines 
indicate the end of 
the bound spectra.}
\label{fig3}
\end{figure*}

In Fig.~\ref{fig3} the sum of the single-particle energies (solid line) is 
compared with the new particle-number smoothed energy (dashed line) and the old
Strutinsky energy (dotted line). The single particle spectrum is the one 
evaluated for the spherical Saxon-Woods mean-field potential for the 
$^{208}$Pb nucleus with the
parameters taken from Ref.~\cite{Ch67}. The background energy $\overline{E}$ is
subtracted from the all three energies presented separately for protons (l.h.s.
plot) and neutrons (r.h.s. plot). One notices that the Strutinsky energy is
always smaller than the present estimate of the smoothed energy. The difference
between both estimates grows with the number of particles and for
$^{208}$Pb (arrows in Fig.~\ref{fig3}) is of the order of a few MeV. This
result is similar to the one obtained with the Ivanyuk and Strutinsky method
\cite{IS79,Iv84}, where the smoothed part of the sum of single-particle
energies was approximated by a local polynomial in the ${\cal N}$-space.

%%%%%%%%%%%%%%%%%%%%%%%%%%%%%%%%%%%%%%%%%%%%%%%%%%%%%%%%%%%%%%%%%%%%%%%%%%%%%%%%

\subsection{Strutinsky smoothed energy}
\label{old}

It is worthwhile to remind here the original Strutinsky method of evaluating of
the smooth energy \cite{St66,NT69,BD72} in order to better understand the 
difference between both approaches. The Strutinsky's way of evaluating the
smooth energy consists of two steps. Firstly one finds the smoothed
single-particle level density $\widetilde g(e)$ and determines the corresponding 
average position $\lambda$ of the Fermi level,
assuming the average particle number conservation. Then in the next step one
evaluates the smoothed energy by integrating the product of
the single-particle energy and smooth level density. It means that in this
method the number of particles is conserved only on the average and the
Strutinsky smoothed energy does not correspond exactly to the averaged sum of
the occupied single-particle energies.

In the Strutinsky shell correction method one evaluates the smooth 
single-particle level density $\widetilde g(e)$ by folding the discrete spectrum 
of eigenstates $e_i$
\be
  g(e) = \sum_i \delta (e-e_i) \,\,,
\label{ge}
\ee
with a smoothing function $j_n(e,e^{'})$ of the $n^{th}$ order which is given 
by Eq.~(\ref{jn}). 
The smooth single-particle level density $\widetilde g(e)$ is then given by
\be
\widetilde g(e)= \sum_i j_n\left(\frac{e-e_i}{\gamma_S} 
\right)\,\,.
\label{getilde}
\ee
Taking the 6$^{th}$ order (so called ``{\it curvature correction}'') polynomial 
into account (see Eq.\ (\ref{fcorr})) the smoothing function has the following
form
\be
  j_6(u) = \frac{1}{\gamma_S\sqrt\pi} e^{-u^2}
       (\frac{35}{16}-\frac{35}{8}u^2+\frac{7}{4}u^4-\frac{1}{6}u^6) \,\,.
\label{jot6}
\ee
The smearing parameter $\gamma_S$ in Eqs.~(\ref{getilde}, \ref{jot6}) is the 
width of the Gauss folding function and should be of the order of the energy 
distance between major shells (i.e. $\hbar\omega_0$) in order to wash out 
the shell structure. 

According to Strutinsky \cite{St66} the average of the sum of the energies of
the occupied single particle levels ($E_{\rm Str}$) is given by the integral
\be
E_{\rm Str} = \int\limits_{-\infty}^{\lambda} 2\,e\,\widetilde g(e)\,de \,\,,
\label{Str1}
\ee
where $\lambda$ is the position of the Fermi energy in the system with 
the washed out shell structure and is fixed by the particle number condition
\be
{\cal N} = \int\limits_{-\infty}^{\lambda} 2\,\widetilde g(e)\,de \,\,.
\label{Str2}
\ee
Here the average number of particles ${\cal N}=Z$ for protons or ${\cal N}=N$
for neutrons. The factor 2 in the above two equations is due to the spin 
degeneracy of the single particle levels. One solves Eq.~(\ref{Str2}) for 
$\lambda$ by iterations.

The Strutinsky energy $E_{\rm Str}$ (\ref{Str1}) is not equal to the average
of the sum of single-particles energies $\widetilde E$, Eq.\ \ref{Etilde}), but
corresponds to the energy of a system which conserves the number of particles 
only on the average (and not exactly as in Eq.~(\ref{Etilde})).
A comparison of the resulting smoothed energies obtained in both methods 
will be presented below.

%%%%%%%%%%%%%%%%%%%%%%%%%%%%%%%%%%%%%%%%%%%%%%%%%%%%%%%%%%%%%%%%%%%%%%%%%%%%%%%%
%%%%%%%%%%%%%%%%%%%%%%%%%%%%%%%%%%%%%%%%%%%%%%%%%%%%%%%%%%%%%%%%%%%%%%%%%%%%%%%%
%%%%%%%%%%%%%%%%%%%%%%%%%%%%%%%%%%%%%%%%%%%%%%%%%%%%%%%%%%%%%%%%%%%%%%%%%%%%%%%%

\section{Comparison of the both estimates of the smoothed energy.}

A significant difference between the new estimate of the 
smooth energy $\widetilde E$ given by Eq.\ (\ref{Etilde}) and the
Strutinsky energy, $E_{\rm Str}$, Eq.\ (\ref{Str1}), is demonstrated in Fig.\ 
\ref{fig3}. It can be easy explained as follows~:

The sum of single-particle energies can be roughly approximated by
Eq.~(\ref{SHO}) as shown in Sec.~\ref{HO}
\be
\overline{\left(\sum\limits_{i=1}^{{\cal N}} e_i\right)} \approx 
              a\,{\cal N}^{4/3} + b\, {\cal N} \,\,,
\label{trend}
\ee
where the parameter $a$ is proportional to the distance $\hbar\omega_0$ 
between major shells and $b$ to the effective depth of the mean-field
potential. Let us assume, just as a matter of discussing our method, that this
average trend represents the true energy sum and that we are dealing with the
degenerate spectrum. Let also ${\cal N}_k$ be the number of particles which can
be placed on the single-particle levels which are below the $k$-th degenerate
level. Note, that the numbers ${\cal N}_k$ (with $k=1,2,...$) are simply the
magic numbers in case of spherical nuclei.

The Strutinsky prescription for the smoothed energy corresponding to 
\be
{\cal N}= \frac{1}{2}({\cal N}_{k+1} - {\cal N}_k)
\ee
particles (i.e. half filled shell) can be written as 
\be
E_{\rm Str} = a{\cal N}^{4/3} + b {\cal N} \,\,,
\ee
while the average over the interval 
$[ {\cal N}_k,{\cal N}_{k+1} ]$ of the energy, Eq.\ (\ref{trend}) 
is given by the integral
\be
\widetilde E = \frac{1}{\Delta} \int\limits_{{\cal N}-\Delta/2}^{{\cal N}+\Delta/2}
           (a n^{4/3} + b n)\, dn \,\,,
\ee
where $\Delta = {\cal N}_{k+1} - {\cal N}_k$ is the degeneracy of the
corresponding single-particle level. In the approximation (\ref{trend}) the 
difference between the average of the single-particle energy sum and the 
Strutinsky energy is
\be
\Delta E = \widetilde E - E_{\rm Str} \approx \frac{1}{54}\, 
         a\, \frac{\Delta^2}{{\cal N}^{2/3}} \,\,,         
\label{diff}
\ee
where $a\approx 3^{4/3}/4\,\hbar\omega_0$.
This approximate expression indicates that the difference between the new 
and the old Strutinsky energy is negligible (of the order of 0.01 MeV for 
heavier nuclei) when the
degeneracy is $\Delta=2$ which is the case for deformed nuclei, while it grows
significantly (up to a couple of MeVs) when the degeneracy is important, as e.g.
in spherical nuclei.

\begin{figure}[h]
  \begin{center}
  \includegraphics[height=8.5cm, angle=270]{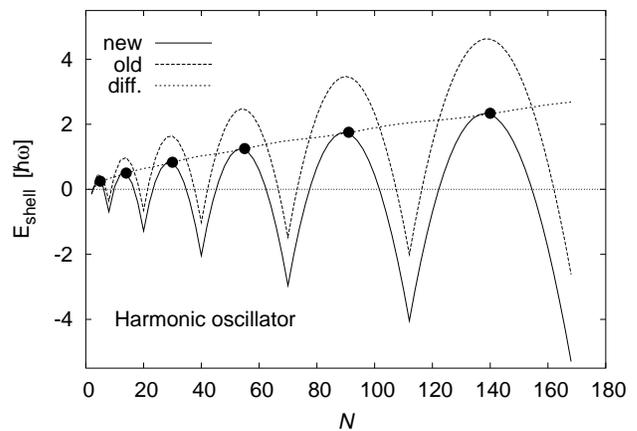}
  \end{center}
\caption{Strutinsky shell energies for a spherically symmetric harmonic 
oscillator potential obtained with the new (Eq.~\ref{Etilde}, solid line) and 
the old (Eq.~\ref{Str1}, dashed line) method (see text) as well as the 
difference between them (dotted line) as function of number of particles 
${\cal N}$. The estimates for the differences $\Delta E$, given by 
Eq.~(\ref{diff}) are marked as the full circles.}
\label{fig4}
\end{figure}

In Fig.~\ref{fig4} we show the shell energy $E_{\rm shell}$ for the spherical
harmonic oscillator single-particle levels in function of the nucleon number
(${\cal N}$). The solid line corresponds to the new approach described in
Sec.~\ref{new} while the dashed one to the old Strutinsky method (see
Sec.~ \ref{old}). The difference between the both estimates and its
approximation with  Eq.~(\ref{diff}) are shown by the dotted line and the full
circles respectively.
\begin{figure*}
  \begin{center}
  \includegraphics[height=17cm, angle=270]{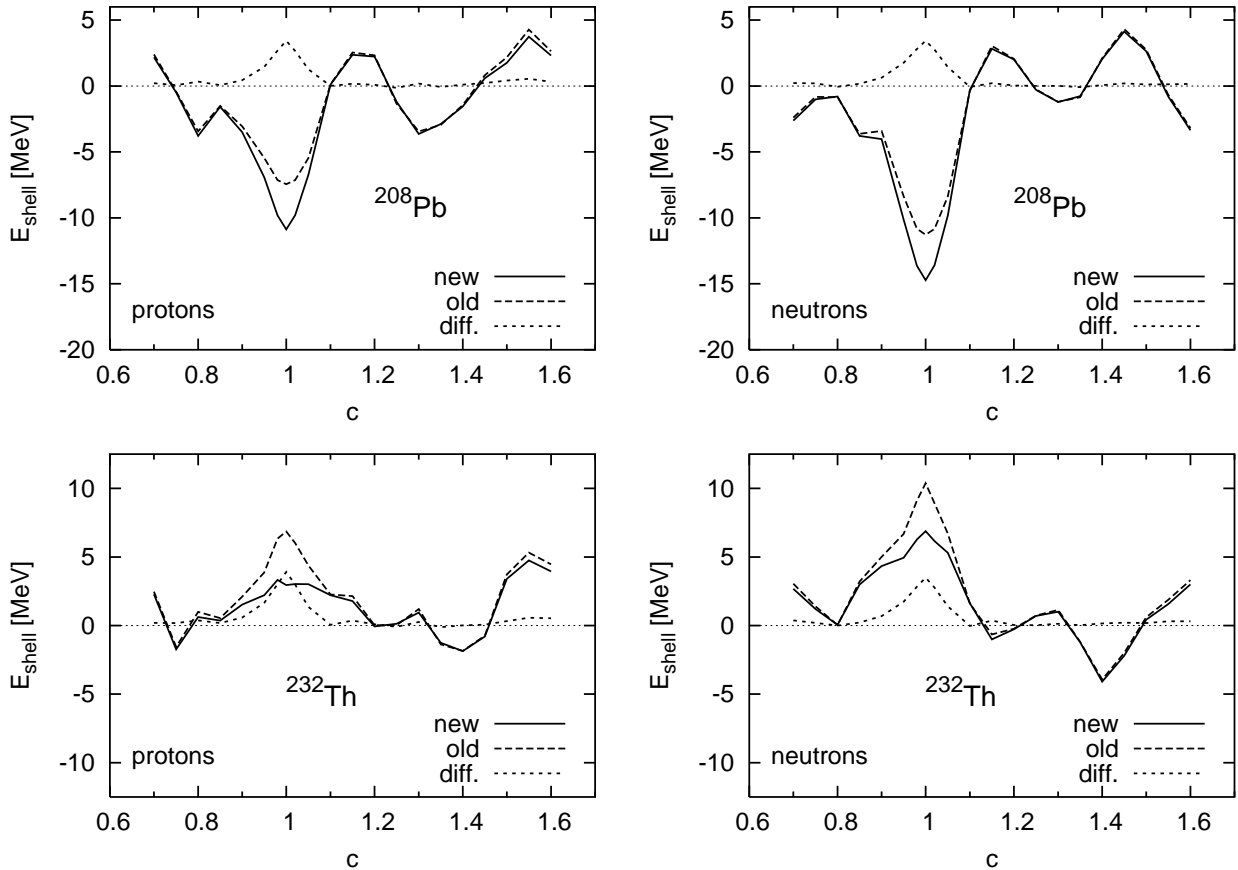}
  \end{center}
\caption{Proton (left) and neutrons (right) Strutinsky shell energies  for the
Saxon-Woods single-particle levels of $^{208}$Pb (top) and $^{232}$Th  (bottom)
obtained with the new (Eq.~\ref{Etilde}, solid line) and the old
(Eq.~\ref{Str1}, dashed-line) method as well as their difference (dotted line)
as functions of the elongation parameter $c$ \protect\cite{BD72}.}
\label{fig5}
\end{figure*}

The proton and neutron shell energies $E_{\rm shell}$ for the nuclei
$^{208}$Pb  and $^{232}$Th obtained with a deformed Saxon-Woods potential are
plotted in  Fig.~\ref{fig5} as function of the elongation parameter $c$ of
Ref.\ \cite{BD72}.  The solid line corresponds to the new approach described in
Sec.~\ref{new}  while the dashed one to the old Strutinsky method (see
Sec.~\ref{old}). The difference between the both estimates is shown by the
dotted line.  The parameters of the Saxon-Woods potential are taken from
Ref.~(\cite{Ch67}).  It is seen that the difference between the shell energies
evaluated with the new particle number conserving method and the traditional
Strutinsky approach becomes negligible with growing nuclear deformation, i.e.
when the degeneracy of single particle levels is lifted, a result which is in
line with the prediction of the approximate expression (\ref{diff}).

\begin{figure*}
  \begin{center}
  \includegraphics[height=17cm, angle=270]{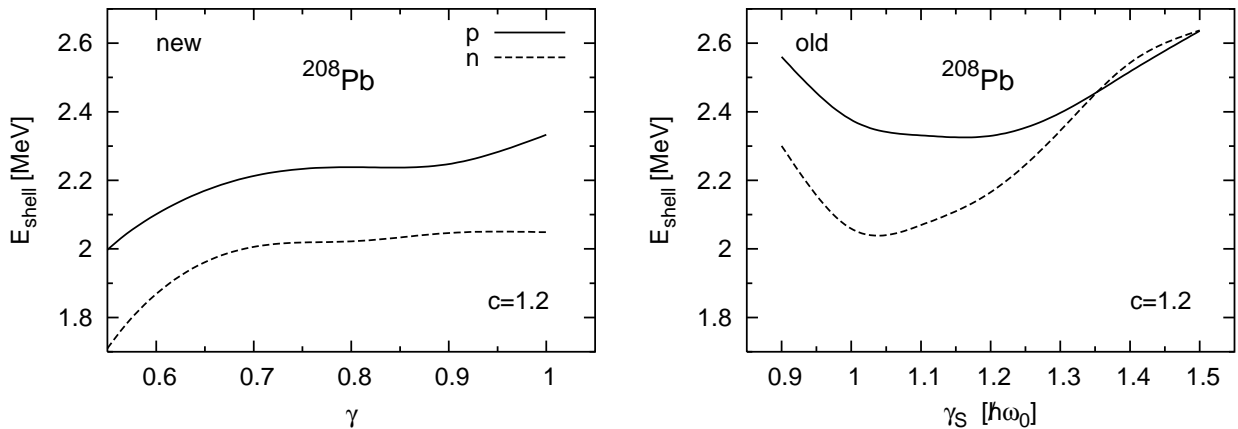}
  \end{center}
\caption{Dependence of proton (solid line) and neutron (dashed line)
Strutinsky shell energies as function of the smearing parameter $\gamma$ (or
$\gamma_S$) obtained with the new (Eq.~(\ref{Etilde}), left) and the old
(Eq.~(\ref{Str1}), right) estimates for the smooth energy for a nucleus
$^{208}$Pb described by a appropriate  Saxon-Woods potential at a deformation
of $c=1.2$.}
\label{fig6}
\end{figure*}

In both (new and old) Strutinsky shell correction methods it is very important 
to choose the appropriate value of the smearing parameter ($\gamma$ or 
$\gamma_S$). This is usually done by fixing its value such that the obtained 
shell energy is, over a certain range of that parameter independent 
of its specific value, a constraint known as the ``{\it plateau condition}'' 
of the Strutinsky method. 
Typical examples of such plateaus are shown in Fig.~\ref{fig6}, where the shell 
energies for $^{208}$Pb at a deformation of $c=1.2$ obtained with the 
appropriately chosen Saxon-Woods potential \cite{Ch67} are drawn. It is seen in
Fig.~\ref{fig6} that our new approach leads to a much more pronounced plateau 
as compared to the old method. Note the different values of the smoothing 
parameter ($\gamma \approx 0.78$ for  both kinds of particles versus $\gamma_S 
\approx 1.20 \hbar\omega_0$ for protons and $\gamma_S \approx 1.05 \hbar
\omega_0$ for neutrons).
\begin{figure*}
  \begin{center}
  \includegraphics[height=17cm, angle=270]{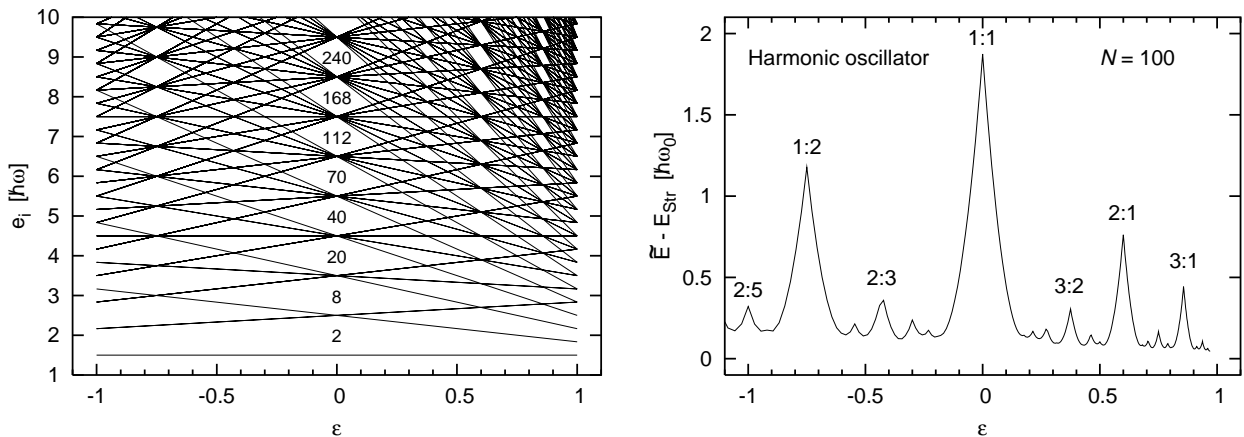}
  \end{center}
\caption{The eigenfunction of the deformed harmonic oscillator potential (l.h.s.
plot) and the difference between the new ($\widetilde E$, Eq.~(\ref{Etilde})) 
and the old ($E_{Str}$, Eq.~(\ref{Str1})) estimates of the smooth energy as 
function of the axial quadrupole deformation $\varepsilon$ \protect\cite{NT69}.}
\label{fig7}
\end{figure*}

Looking at Fig.~\ref{fig5} one could have the impression that a significant
difference between the shell energies obtained by the averaging in the 
particle-number or single-particle energy spaces appears around spherical
shapes and  vanishes with growing deformation.  On the other hand,
Eq.~(\ref{diff}) predicts that a significant difference between both types of
averaging procedures should appear whenever a large degeneracy of the
single-particle levels is present. A good example of the spectrum which becomes
strongly degenerate at some deformation  points are the eigenenergies of the
anisotropic harmonic oscillator.  The degeneracy clearly appears at those
ellipsoidal deformation ($\varepsilon$, \cite{NT69}) points where the ratio of
the axes is equal to the ratio of small integers as can be seen in the
left part of Fig.~\ref{fig7}, where such a spectrum is plotted. The harmonic
oscillator single-particle levels presented there were used to  evaluate the
shell energy of a system composed of ${\cal N}$=100 particles. The difference
between the new and the old estimates  ($\widetilde E$ and $E_{\rm Str}$) is
plotted as function of the  quadrupole deformation parameter
$\varepsilon$ \protect\cite{NT69}. A large difference between both approaches
appears where the degeneracy of levels  grows. A similar effect was already
observed in Ref.~\cite{IS79}. This result  clearly shows that using the new
approach one can expect some modifications in  the potential energy surface not
only around spherical shapes but also at deformations which correspond to the
super- or hyper-deformed isomers.

%%%%%%%%%%%%%%%%%%%%%%%%%%%%%%%%%%%%%%%%%%%%%%%%%%%%%%%%%%%%%%%%%%%%%%%%%%%%%%%%
%%%%%%%%%%%%%%%%%%%%%%%%%%%%%%%%%%%%%%%%%%%%%%%%%%%%%%%%%%%%%%%%%%%%%%%%%%%%%%%%
%%%%%%%%%%%%%%%%%%%%%%%%%%%%%%%%%%%%%%%%%%%%%%%%%%%%%%%%%%%%%%%%%%%%%%%%%%%%%%%%

\section{Summary and conclusions}

A new method of evaluating the smooth part of the total single-particle energy
is proposed. The folding of the sums of single-particle energies is performed
in the particle-number space rather than in the one of single-particle energies
as done in the old Strutinsky method. The averaging in the  ${\cal N}$-space is
consistent with the definition of the macroscopic energy component which
represents the average behavior in $Z$ and $A$ of the nuclear binding energy.

One has also to notice that the integral over ${\cal N}$ of the shell energy 
evaluated with the new prescription is close to zero for sufficiently large
number of particles, while this is not the case for the traditional Strutinsky 
shell correction which grows systematically with ${\cal N}$. This indicates
that the original Strutinsky prescription generates shell energy which do not
fluctuate around zero when the number of particles is increased which modifies
systematically the macroscopic part of the energy. In addition this
modification depends on the shape of nucleus, so that this deficiency of the
traditional approach cannot be corrected by an adjustment of the parameters of
the macroscopic energy.

The new estimate differs significantly from the Strutinsky smoothed energy when
a large degeneracy of the single-particle levels is present as this is the
case in spherical and nearly-spherical nuclei or in some shape isomers. In
such nuclei the shell energy is shifted down by a few MeV with respect the old
predictions while its amplitude is almost unchanged as function of the particle
number. This means that the macroscopic-microscopic method with the new 
estimate of the shell energy will, {\it when leaving the parameters of the
macroscopic and microscopic parts untouched}, predict the magic and quasi-magic
nuclei as well as some shape isomers more bound than this was predicted by the
calculations done with the old Strutinsky method. Also the deformation energies
of non-magic nuclei will be smaller and one could obtain different equilibrium
shapes (i.e. ground state quadrupole moments). The fission barrier for
spherical nuclei will be also significantly increased and, as a consequence,
the spontaneous fission of such nuclei (e.g. some super-heavy isotopes) will be
less probable. In addition the $Q$-value for an $\alpha$-decay will be
modified when it occurs between deformed and spherical isotopes (or vice
versa). This means that the consequences of the naive use of new method could
be dramatic. 

I would therefore like to end with the following warning:\\  {\it Do not use
the new prescription for the shell energy (Eqs. (\ref{Sn}) - (\ref{Etilde})) in
practical calculations without an appropriate readjusting of the parameters of
the mean-field potentials (e.g. Saxon-Woods, Nilsson, Yukawa-folded),
macroscopic models (e.g. liquid drop, finite range droplet or Thomas-Fermi),
and the pairing force.}

%%%%%%%%%%%%%%%%%%%%%%%%%%%%%%%%%%%%%%%%%%%%%%%%%%%%%%%%%%%%%%%%%%%%%%%%%%%%%%%%

\acknowledgments

The author wishes to express his thanks for the warm hospitality extended to
him by the Institute for Subatomic Research (IReS) and the Louis Pasteur
University of Strasbourg where a part of this work was done. Discussions
with Professors Jerzy Dudek who was {\it Spiritus Movens} of this research and
Johann Bartel from IReS as well as Professor Fedor Ivanyuk of the Institute of 
Nuclear Physics in Kiev were also very helpful.

%%%%%%%%%%%%%%%%%%%%%%%%%%%%%%%%%%%%%%%%%%%%%%%%%%%%%%%%%%%%%%%%%%%%%%%%%%%%%%%%
%%%%%%%%%%%%%%%%%%%%%%%%%%%%%%%%%%%%%%%%%%%%%%%%%%%%%%%%%%%%%%%%%%%%%%%%%%%%%%%%
%%%%%%%%%%%%%%%%%%%%%%%%%%%%%%%%%%%%%%%%%%%%%%%%%%%%%%%%%%%%%%%%%%%%%%%%%%%%%%%%

\appendix

\section{Folding of discrete data}

%%%%%%%%%%%%%%%%%%%%%%%%%%%%%%%%%%%%%%%%%%%%%%%%%%%%%%%%%%%%%%%%%%%%%%%%%%%%%%%%

\subsection{General formulae}

Our aim is to approximate a sample of $N$ ordered points $\{x_i,y_i\}$ by a
continues smooth function $\widetilde y(x)$. We would like to solve this
problem using the Gauss-Hermite folding method which was idea the originally
proposed by V.M. Strutinsky \cite{St66} and later-on generalized in
Ref.~\cite{NT69}. Having the width of the folding function comparable with the
average distance between points $x_i$ one can obtain the folded function which
is very close to the data points but increasing this width one can also wash
out the fine structure stored in the data. Usually the Strutinsky method was
used to realize the second scope. The parameter of the folding procedure will
be determined by requirement that the integral of the folded function should be
the same as the integral evaluated with the sample of $\{x_i,y_i\}$ pairs using
the trapezium  rule. 

Let $j_{n}(x,x')$ be a symmetric function of its arguments 
(i.e. $j_{n}(x,x')=j_{n}(x',x)$) having the following properties:
\be
  \int\limits_{-\infty}^{+\infty} j_{n}(x,x')\,dx = 1 
\label{norm}
\ee
and
\be
  P_{k}(x) = \int\limits_{-\infty}^{+\infty} P_{k}(x')\,
                    j_{n}(x,x')\, dx' \,\,,
\label{Strut}
\ee
where $k \leq n$ are even natural numbers and $P_{k}(x)$ is an arbitrary 
polynomial of order $k$. In the following, the function $j_n(x,x')$ will  be
called the folding function of the $n^{\rm th}$ order. The last equation ,
frequently called the {\it Strutinsky condition} ensures that the folding does
not change the average behavior of the function $Y(x)$ which is represented by
the ensemble of $\{x_i,y_i\}$ points. An example of such a folding function can
be a combination of the Gauss function and the Hermite polynomials of the
argument proportional to $|x-x'|$, frequently used in the  Strutinsky shell
correction method \cite{St66,NT69}. More  detailed description of  such a
folding function will be given in the next section.

With each discrete point $(x_{i},y_{i})$ one can associate the function 
$\widetilde{y}_{i}(x)$ defined by:
\be
\widetilde{y}_{i}(x) = \int\limits_{-\infty}^{+\infty} y_{i}\, \delta(x' - x_{i})\,
                  j_{n}(x,x')\, d x' \,\,,
\ee
where $\delta(x)$ is the Dirac $\delta$-function. A straightforward 
calculation gives
\be
  \widetilde{y}_{i}(x) =  y_{i}\, j_{n} (x,x_{i}) \,\,.
\ee
Using Eq.~(\ref{norm}) it is easy to verify that the integral of the function 
$\widetilde{y}_{i}(x)$ is
\be
\int\limits_{-\infty}^{+\infty}\widetilde{y}_{i}(x)\,dx = y_{i} \,\,.
\ee
Let us construct the function $\widetilde{y}(x)$ by summing up, with weight $w_i$
all functions $\widetilde{y}_{i}(x)$
\be
  \widetilde{y}(x) = \sum_{i=1}^{N} w_{i}\, \widetilde{y}_{i}(x) \,\,.
\ee
The function $\widetilde{y}(x)$ is an approximation of $y(x)$ if the weights
$w_{i}$ are determined from the assumption that the integrals of the unfolded
and folded function are (nearly) equal:
\be
  \sum_{i=1}^{N}  y(x_{i})\,\Delta x_{i} = \int\limits_{-\infty}^{+\infty}
              \widetilde{y}(x)\,dx =  \sum_{i=1}^{N} w_{i}\, y_{i} \,\,,
\label{int}
\ee
where $\Delta x_{i}$ is set to: 
\be
  \Delta x_{i} = \frac{1}{2}\,(x_{i+1} - x_{i-1})  \,\,.
\ee
Eq. (\ref{int}) implies that a reasonable choice of the weight is 
\be
w_{i} = \Delta x_{i} \,\,.
\ee
Thus the folded function $\widetilde y(x)$ is given by
\be
\widetilde y(x) = \sum_{i=1}^N  y_i\,\Delta x_i\, j_n(x,x_i) \,\,.
\label{appy}
\ee

%%%%%%%%%%%%%%%%%%%%%%%%%%%%%%%%%%%%%%%%%%%%%%%%%%%%%%%%%%%%%%%%%%%%%%%%%%%%%%

\subsection{Gauss-Hermite folding function}

Let the folding function $j_n(x,x')$ be defined with the help of the Gauss 
function as
\be
 j_n(x,x') = \frac{1} {\gamma\sqrt{\pi}} \,
             {\rm exp}\left[-\left(\frac{x-x'}{\gamma}\right)^2\right]
             f_n\left(\frac{x-x'} {\gamma}\right)\,, 
\label{jn}
\ee
where $\gamma$ is a parameter and $f_n(\frac{x-x'}{\gamma})$ is the so called
corrective polynomial of $n^{th}$ order, determined by the Strutinsky
condition  (\ref{Strut}). In the following we would like to evaluate the
coefficients of the corrective  polynomial using some properties of the Hermite
polynomials which are orthogonal with the weight equal to the Gauss function.

Let us introduce variable $u = (x - x')/\gamma$ defined in the
interval $(-\infty ,+\infty )$. The smearing function $j_n(x,x')$ and the
polynomial $P_n(x)$ in (\ref{Strut}) can now be written as
\be 
 j_n(x,x') = \frac{{\rm e}^{-u^2}} {\gamma\sqrt{\pi}} \,f_n(u) \,,
\ee

\be
  P_n(x') =  P_n(x - \gamma\, u) \equiv   {P_n}'(u)\,, 
\ee
and
\be
  P_n(x) =  P_n(x + \gamma\, 0) \equiv   {P_n}'(0)\,.
\ee
The so far arbitrary polynomial ${P_n}'(u)$ can be written down as a series of 
the Hermite polynomials of order $i$
\be
 {P_n}'(u) = \sum^{n }_{i=1}a_{i}\,H_{i}(u)\,\,.
\label{hep}
\ee
Now the condition (\ref{Strut}) can be written as
\be
 {P_n}'(0) = \frac{1}{\sqrt{\pi}} \int\limits^{+\infty }_{-\infty } \,
 {P_n}'(u){\rm e}^{-u^{2}} \,f_n(u)\, du 
\label{Pn0}
\ee
and inserting relation (\ref{hep}) into (\ref{Pn0}) one obtains 
\be
 \sum^{\rm p}_{i=1}a_{i} \left\{\frac{1}{\sqrt{\pi}} \int^{+\infty}_{-\infty}
  {\rm e}^{-u^{2}} \, H_{i}(u) \, f_n(u)\,du - H_{i}(0)\right\} = 0 \,\,.
\ee
On the other hand, the last equation should be fullfiled for arbitrary values 
of $a_i \neq 0$) what leads to the following set of equations
\be
 \frac{1}{\sqrt{\pi}} \int^{+\infty}_{-\infty} \, {\rm e}^{-u^2} \, H_{i}(u) \,
         f_n(u)\,du = H_{i}(0)  \,\,,
\label{coef}
\ee
where  $i=0,2,...,n$.
From the other side the corrective function $f_n(u)$ can be also decomposed 
in terms of the Hermite polynomials
\be
 f_n(u) = \sum^{n}_{k=1}C_{k}\,H_{k}(u)\,\,.
\label{corr}
\ee
Inserting the above relation into Eq.~(\ref{coef}) gives
\be
H_i(0) = \sum^{r}_{k=1}C_{k} \frac{1}{\sqrt{\pi}} \int^{+\infty}_{-\infty} \,
   {\rm e}^{-u^{2}}H_{i}(u) \, H_{k}(u)\,du \,\,.
\label{rown}
\ee
Then using the orthogonality properties of the Hermite polynomials
\be
\frac{1}{\sqrt{\pi}}\int^{+\infty}_{-\infty} \, {\rm e}^{-u^{2}}H_{i}(u) \, 
         H_{k}(u)\,du = 2^{i}\,i!\,\delta_{ik} \,\,,
\ee
one obtains the coefficients of the corrective polynomial (\ref{corr})
\be
  C_{i} = \frac{1}{2^{i}i!}\,H_{i}(0)
\ee
The values of the Hermite polynomials at zero are
\be
  H_i(0) = \left\{2^n
 \begin{array}{cl}
         1            & \mbox{for}~i=0        \,\,,\\
  (-1)^n (2n-1)!!~~~~ & \mbox{for}~i = 2n     \,\,, \\
       0              & \mbox{for}~i = 2n + 1 \,\,,
\end{array}\right.
\label{zero}
\ee
so that
\be
 C_i = \left\{
 \begin{array}{cl}
           1                         & \mbox{for}~i=0         \,\,,\\
(-1)^n \frac{(2n-1)!!}{2^n(2n)!}~~~~ & \mbox{for}~i = 2n > 0  \,\,,\\
           0                         & \mbox{for}~i = 2n + 1  \,\,.
\end{array}\right.
\label{Ci}
\ee

The first few coefficients $C_i$ and the corresponding Hermite polynomials 
are:\\
\be
\begin{array}{lll}
C_0 = 1		    &,	         &  H_0(u) = 1 \,\,,		        \\
C_2 = - \frac{1}{4}  &,		 &  H_2(u) =  4u^2 - 2 \,\,,	        \\
C_4 = + \frac{1}{32} &,		 &  H_4(u) = 16u^4 -  48u^2 + 12 \,\,,  \\
C_6 = - \frac{1}{384}&,&   H_6(u) = 64u^6 - 480u^4 + 720u^2 - 120 \,\,,  \\ 
\end{array}
\ee
and the resulting corrective polynomials have the following form
\be
\begin{array}{l} 
  f_0(u) = 1                                                          \,\,, \\
  f_2(u) = \frac{3}{2}  -              u^2                            \,\,, \\
  f_4(u) = \frac{15}{8} -  \frac{5}{2} u^2 + \frac{1}{2} u^4          \,\,, \\ 
  f_6(u) = \frac{35}{16}- \frac{35}{8} u^2 + \frac{7}{4} u^4 - \frac{1}{6}u^6
          \,\,, \\
\end{array}
\label{fcorr}
\ee

Finally the function $\widetilde y(x)$ approximated using the Gauss-Hermite 
folding reads:
\be
\widetilde y(x) = \frac {1}{\gamma\sqrt{\pi}}\,\sum_{i=1}^N  y_i\,\Delta x_i \,
    {\rm exp}\left[-\left(\frac{x-x_i}{\gamma}\right)^2\right]
    \,f_n\left(\frac{x-x_i}{\gamma}\right) \,\,.
\label{GHF}
\ee
In principle the smearing parameter $\gamma$ is arbitrary and it can be
different at each point $x_i$. But it should be related to the distance $\Delta
x_i$ between subsequent points if one would like to approximate the function
stored in the mesh of $\{x_i,y_i\}$ points. Similarly one has to choose $\gamma$
of the order of the period-length of the fine structure (e.g. shell effects) 
in case when one would like to wash out this structure from the function $y(x)$.

%%%%%%%%%%%%%%%%%%%%%%%%%%%%%%%%%%%%%%%%%%%%%%%%%%%%%%%%%%%%%%%%%%%%%%%%%%%%%%%%
%%%%%%%%%%%%%%%%%%%%%%%%%%%%%%%%%%%%%%%%%%%%%%%%%%%%%%%%%%%%%%%%%%%%%%%%%%%%%%%%
%%%%%%%%%%%%%%%%%%%%%%%%%%%%%%%%%%%%%%%%%%%%%%%%%%%%%%%%%%%%%%%%%%%%%%%%%%%%%%%%


\begin{thebibliography}{10}
\bibitem{St66} V.M. Strutinsky, Sov. J. Nucl. Phys. {\bf 3}, 449 (1966);
               Nucl. Phys. {A95}, 420 (1967); Nucl. Phys. {\bf A122}, 1 (1968).
\bibitem{MS66} W.D. Myers, W.J. \'Swi\c{a}tecki, Nucl. Phys. {\bf 81}, 1 (1966).
\bibitem{NT69} S.G. Nilsson, C.F. Tsang, A. Sobiczewski, Z. Szyma\'nski,  
               S. Wycech, S. Gustafson, I.L. Lamm, P. M$\ddot{o}$ller, 
               B. Nillson, Nucl. Phys. {\bf A131}, 1 (1969).
\bibitem{BD72} M. Brack, J. Damgaard, A.S. Jensen, H.C. Pauli, V.M. Strutinsky, 
               C.Y.Wong, Rev.\ Mod.\ Phys.\ {\bf 44}, 320 (1972).
\bibitem{SI75} V.M. Strutinsky, F.A. Ivanyuk, Nucl. Phys. {\bf A255}, 405 
               (1975).
\bibitem{SI77} V.M. Strutinsky, F.A. Ivanyuk, Izvestia AN SSSR, {\bf 41}, 114
               (1977).
\bibitem{IS78} F.A. Ivanyuk, V.M. Strutinsky, Z. Phys. {\bf A286}, 291 (1978);
               Z. Phys. {\bf A290}, 107 (1979).
\bibitem{IS79} F.A. Ivanyuk, V.M. Strutinsky, Z. Phys. {\bf A293}, 337 (1979).
\bibitem{St79} V.M.Strutinsky, {\it Shell structure in fission}, Proc. Symp.
               on "Physics and Chemistry of Fission 1979", Julich, 14-18 May 
               1979, vol. 1, p.475-500, IAEA, Vienna, 1980.
\bibitem{Iv84} F.A. Ivanyuk, Z. Phys. {\bf A316}, 233 (1984).
\bibitem{PD03} K. Pomorski, J. Dudek, Phys. Rev. {\bf C67}, 044316 (2003).
\bibitem{SG77} A. Sobiczewski, A. Gyurkovich, M. Brack, Nucl. Phys. {\bf A289},
               346 (1977).
\bibitem{BM69} A. Bohr, B.R. Mottelson, {\it Nuclear Structure},vol. 1 and 2, 
               W.A. Benjamin Inc, New York, 1969 and 1974.
\bibitem{BG84} J. F. Berger, M. Girod and D. Gogny, Nucl. Phys. {\bf A428},
               236 (1984).
\bibitem{Ch67} A. Chepurnov, Yad. Fiz. {\bf 6}, 955 (1967).
\end{thebibliography}
\end{document}